\documentclass{elsart}
\usepackage[T1]{fontenc}
\usepackage[latin1]{inputenc}
\usepackage{graphicx}
\usepackage{setspace}

\makeatletter

\providecommand{\LyX}{L\kern-.1667em\lower.25em\hbox{Y}\kern-.125emX\@}

\makeatother
\begin{document}
\begin{frontmatter}
\title{Chaotic Jets}

\author[label1]{Xavier Leoncini\corauthref{cor1}} and
\ead{leoncini@up.univ-mrs.fr}
\author[label2,label3]{George M. Zaslavsky}
\ead{zaslav@cims.nyu.edu}
\corauth[cor1]{Corresponding author.}

\address[label1]{LPIIM, Equipe Dynamique des Syst\`emes complexes, CNRS, Universit\'e de Provence, Centre Universitaire de Saint J\'er\^ome, F-13397 Marseilles, France}
\address[label2]{ Courant Institute of Mathematical Sciences, New York University, 251
Mercer St., New York, NY 10012, USA}
\address[label3]{Department of Physics, New York University, 2-4 Washington Place, New
York, NY 10003, USA}

\begin{abstract}
The problem of characterizing the origin of the non-Gaussian properties
of transport resulting from Hamiltonian dynamics is addressed. For
this purpose the notion of chaotic jet is revisited and leads to the
definition of a diagnostic able to capture some singular properties
of the dynamics. This diagnostic is applied successfully to the problem
of advection of passive tracers in a flow generated by point vortices.
We present and discuss this diagnostic as a result of which clues
on the origin of anomalous transport in these systems emerge.
\end{abstract}
\end{frontmatter}

\section{Introduction}

The characterization of the kinetics emerging from Hamiltonian dynamics
has been a long going problem which dates back to Boltzmann. As the
literature evolves, more and more Hamiltonian systems are showing
anomalous properties, in the sense that their kinetics does not follow
a simple Gaussian process but rather gives rise to what one now calls
{}``strange kinetics'' \cite{Schlesinger93,Zaslavsky2002}. In systems which belong
to the of $3/2$ degree of freedom Hamiltonians, the origin of these
anomalous properties can be relatively well understood when a portrait
of the phase space using Poincaré maps is drawn. The system is not
ergodic, and a well-defined stochastic sea filled with various islands
of regular motion is observed. The anomalous properties and their
multi-fractal nature are then linked to the existence of islands within
the stochastic sea and the phenomenon of stickiness observed around
them. 

One of such systems corresponds to the problem of the advection of
passive particles in flows generated by three vortices. \cite{Kuznetsov98,Kuznetsov2000,LKZ01}.
Special islands also known as {}``vortex cores'' are surrounding
each of the three vortices. Transport in these systems is anomalous,
and the exponent characterizing the second moment exhibit a universal
value close to 3/2. Special interest in these point vortex systems
follows from different observations and models that have exhibited
anomalous transport properties\cite{Chernikov90,Solomon94,Weeks96,Kovalyov2000,Provenzale99,Hansen98}.
The applications to geophysical flows where advected quantities vary
from the ozone in the stratosphere to various pollutants coupled with
the rise of the environmental concerns, make the understanding of
these anomalous properties even more crucial. Systems governed by
three point vortices give rise to the phenomenon know as chaotic advection,
the quasi-periodic flow allows the existence of Lagrangian chaos which
enhances considerably the mixing properties. The non-uniformity of
the phase space and the presence of islands have a considerable impact
on the transport properties. The phenomenon of stickiness on the boundaries
of the islands generates strong {}``memory effects'' as a result
of which transport becomes anomalous. However, when the flows is itself
chaotic and one cannot easily draw a phase portrait, it is crucial
to define a proper diagnostic which will be able to capture some singular
properties of the dynamics that would give clues on the origin of
the anomalous transport. %
\begin{figure}
\begin{center}\includegraphics[  width=7cm]{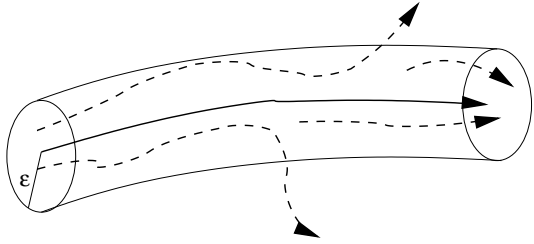}\end{center}

\caption{Tracking of $\epsilon $ coarse-grained regular jet.\label{cap:Tracking-of-jet}}
\end{figure}

There have been already different attempt to create a tool able to
track this phenomenon \cite{Boatto99,Castiglione99,Andersen2000}.
Most of them originates from the definition of the Lyapunov exponent
\begin{equation}
\sigma _{L}=\lim _{r(0)\rightarrow 0}\lim _{\tau \rightarrow \infty }\frac{1}{\tau }\ln \frac{r(\tau )}{r_{0}}\: ,\label{lyapunovdefinition}\end{equation}
where $r_{0}$ is the initial separation between two nearby trajectories
and $r(\tau )$ is the separation at time $\tau $. When the dynamics
is ergodic this exponent gives a global (non-local) signature of the
degree of chaoticity of the system. For instance, Finite-Time Lyapunov
exponents (FTLE) are used. These exponents are measured from trajectories
whose initial conditions are covering the plane, resulting in a scalar
field distributed within the space of initial conditions. Regions
of vanishing FTLE can then be identified as sources of {}``long memory''
effects \cite{Boatto99}. The difficulties with these types of approach
resides in the introduction of arbitrary choice free parameters when
computing FTLE, namely the initial separation between two different
trajectories $r_{0}$ and the time interval $\tau $ within which
they are computed: more or less arbitrary specific scales are set
both in time and space. It is likely that $r(t)$ is not always smooth
growing function of time on the scale of an arbitrary time $\tau $
and jumps between different spatial scales, each with a potential
physical meaning. We can anticipate that this may be especially the
case when different regions of small (if not zero) Lyapunov exponents
are present in the system. Moreover since the field of FTLE is computed
within the initial conditions space, this method is rather numerically
expensive if one wishes to follow the regions of vanishing FTLE through
time. 

In the following we discuss a diagnostic which is greatly inspired
from a natural phenomena typically observed in geophysical flows,
namely the presence of jets. Indeed one can picture a jet as an ensemble
of fluid particles traveling {}``coherently'' together and exhibiting
on a given scale little dispersion. This phenomenon does not discard
the possibility of strong chaotic motion within the jet, but restrict
it within a specific scale. This approach is in fact under certain
aspect very similar to measuring regions of vanishing FTLE but has
the advantage of clearing out some of its shortcomings. In a non ideal
situation, typically when dealing with numerical or experimental data,
we only have a finite spatial resolution and have only access to a
finite portion of a trajectory (finite time). In some sense we are
facing a {}``coarse grained'' phase space, and each point is actually
a ball from which infinitely many real trajectories can depart. Hence,
two nearby real trajectories may diverge exponentially for a while
but then get closer again without actually leaving our resolution
scale, a process which may happen over and over; in this {}``coarse
grained'' perspective those two real trajectories are identical.
It may then possible that within the phase space exist nearby trajectories
which remain within our scale of interest for a relatively large time,
giving rise to what we call a \emph{chaotic jet, or simply jet} \cite{Afanasiev91,LZ02,Afraimovich03,Carreraspreprint}.
In this situation we are typically interested on the chaotic properties
of the system from the resolution scale and up and dismay any chaotic
motion which may occur within the jet. 

We shall now discuss the strategy  implemented to detect
 jets. Let us consider a trajectory $\mathbf{r}(t)$ within the
phase space. We associate to this trajectory its corresponding {}``coarse
grained'' equivalent in the following way: for each time $t$, we
consider a ball $B(\mathbf{r}(t),\epsilon )$ of radius $\epsilon $
whose center is the position $\mathbf{r}(t)$. The $\epsilon $-coarse
grained trajectory is then formed by the reunion of the balls for
all time $\cup B(\mathbf{r}(t),\epsilon )$ and defined by our minimal
scale of interest $\epsilon $. Once this $\epsilon $-coarse grained
trajectory is defined, we look for real trajectories within the ball
at a given time. We then measure two quantities: a time $\tau $ and
length $s$, corresponding to actually how long the trajectory remains
and how much it travels before its first escape from the coarse grained
trajectory (see Fig. \ref{cap:Tracking-of-jet}).

This approach has already been used with success when studying numerically
the advection of passive tracers in flows governed by point vortices
\cite{LZ02}. In this setting, the velocity field generated by the
chaotic motion of four point vortices was considered and the jet data
was collected as follows: given an initial condition of a tracer,
test particles are placed in its neighborhood (typically two), at
a distance $\delta =10^{-6}\ll \epsilon =0.03$. $\epsilon $ has
been chosen to be typically much smaller than the characteristic small
scale of the system, namely the radius of the core surrounding the
vortices, which was estimated around $0.25$. In this setting each
time a test particles reaches a distance $\epsilon =0.03$ (the width
of the coarse grained trajectory), the time interval $\tau $ and
the distance traveled $s$ are recorded, the test particle is then
discarded, a new one takes its place at a distance $\delta $ and
so on. The main difficulty in using this diagnostic follows from the
fact that data acquisition is not sampled linearly in time nor space.
Thus a careful choice for the values of $\epsilon $ and $\delta $
is necessary in order to avoid accumulating too much data while still
capturing the rare events. Once the parameters are defined, we first
look at the trapping time distribution. A power law decay with typical
exponent $\gamma =2.82$ %
\begin{figure}[htbp]
\begin{center}\includegraphics[  width=7cm,
  keepaspectratio]{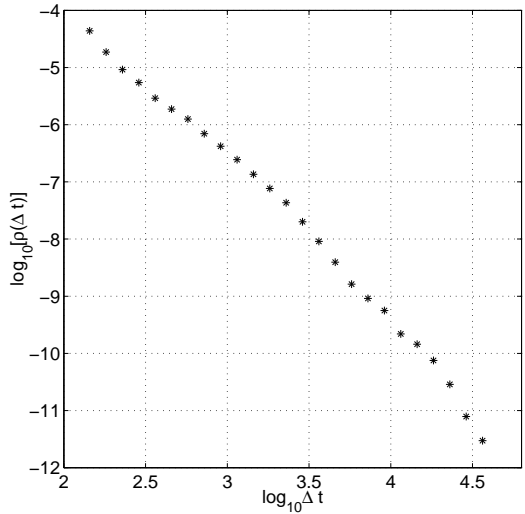}\end{center}

\caption{Tail of the distribution of trapping times $\Delta t$. A power-law
decay, with some oscillations is observed. Typical exponent is $\rho (t)\sim t^{-\gamma }$
with $\gamma \approx 2.823$.\label{Figescapetime4vortex}}
\end{figure}
. is observed in in Fig. \ref{Figescapetime4vortex} for the system
driven by four vortices. The initial condition of the vortex system
is identical as the one used in \cite{Laforgia01}. The data corresponds
to 4 different real trajectories. The time of the simulation is $5.10^{6}$,
the time step is $0.05$, and the evolution of the vortices and passive
tracers was computed using a fifth order Gauss-Legendre simplectic
scheme\cite{McLachlan92}. This power law shows that long lived jets
exists and are responsible for the anomalous transport properties
observed in this system (see \cite{LZ02} for details).

In order to identify dynamically that a reference tracer is located
within a jet it is convenient to go back the definition of the Lyapunov
exponent (\ref{lyapunovdefinition}), which inspires two different
types of FTLE: \begin{equation}
\sigma _{L}=\frac{1}{\Delta t}\ln \frac{\epsilon }{\delta }\: ,\hspace {10mm}\sigma _{D}=\frac{1}{\Delta s}\ln \frac{\epsilon }{\delta }\: ,\label{eq:2}\end{equation}
where contrary to the typical FTLE's the value of the logarithm is
fixed and $\Delta t$ or $\Delta s$ are the variables. These exponent
are very similar to the notion of Finite Size Lyapunov Exponent (FSLE)
introduced in \cite{Aurell97}, however no averages are performed
and the whole distribution is used. %
\begin{figure}
\begin{center}\includegraphics[  width=7cm]{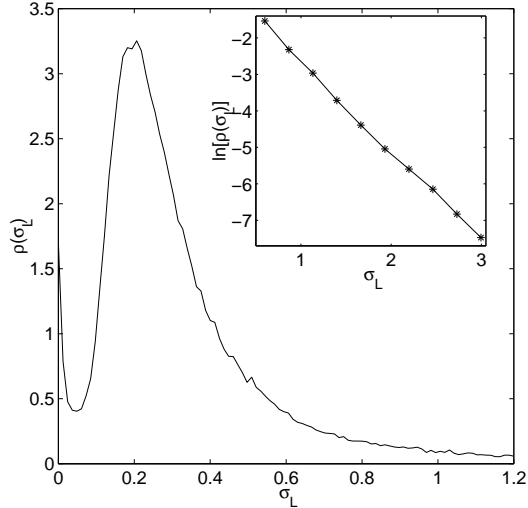}\end{center}

\caption{Distribution of $\sigma _{L}$ (see Eq. (\ref{eq:2})). We notice
an exponential decay for high exponents. $\rho (\sigma _{L})\sim \exp (-\sigma _{L}/\sigma _{L_{0}})$
with $\sigma _{L_{0}}\approx 0.4$. We can notice a minimum around
$\sigma _{L}\approx 0.05$. The observed accumulation near 0 results
from the existence of long lived jets. \label{Figlyapunovtime}}
\end{figure}
 The distribution of the measured $\sigma _{L}$ are illustrated in
Fig. \ref{Figlyapunovtime}. One can notice in this plot two different
types of behavior: For large exponents the distribution decays exponentially
with a characteristic exponent $\sigma _{L_{0}}\approx 0.4$. Since
the speed of tracers is bounded, it will always take a finite time
to escape from the ball, thus an expected maximum value for $\sigma _{L}$.
Regarding the exponential decay behavior before reaching this maximum
value, since we are measuring escape times from a given (moving) region
of the phase space, this exponential behavior can be expected from
a chaotic system with good mixing properties. On the other hand the
existence for the small FTLE's of a local minimum in the probability
density function for a non zero value of $\sigma _{L}$ is more interesting
when dealing when anomalous transport behavior. Indeed, this minimum
characterizes the crossover from the erratic chaotic motion of the
reference tracer to its motion within a regular jet. Indeed if the
tracer is within jet, the test particles are nevertheless expected
to escape from the tracers vicinity but with trapping times exhibiting
a power-law decay, therefore if the passive tracer is evolving within
a jet for a long time, we should expect an accumulations of events
corresponding to test particles leaving the jet. 

The shape of the distribution of $\sigma _{D}$ is qualitatively identical
to the one obtained for $\sigma _{L}$ in Fig. \ref{Figlyapunovtime},
with an exponential decay and a a local minimum $\sigma _{D}*\sim 0.03$
near zero. In fact the measure of $\sigma _{L}$ is biased towards
jets within which the average speed is slow, while by using $\sigma _{D}$
these dynamical differences are erased and only the actual topology
of the vicinity of a trajectory matters. Anomalous transport properties
are most often characterized by measuring the time evolution of the
distribution of some physical coordinate or position, which has the
dimension of a length. In this light it should be clear that the actual
length of jets are going to play an active role the shape of the distribution
of displacements, while the role by the time spent within a jets is
more subtle as it is also dependent on the speed within the jet. Hence
$\sigma _{D}*$ is the adequate parameter to control whether or not
the reference tracer is evolving within a coherent jet, while its
averaged speed $\sigma _{L}/\sigma _{D}$ allows to differentiate
between different types of jets (see \cite{LZ02} for details). 

In order to dynamically look for coherent jets, a tracer's evolution
with its test tracers is studied, once the threshold given by $\sigma _{D}*$
is reached, the measured $\sigma _{D}$ will be such that $\sigma _{D}<\sigma _{D}^{*}$,
hence the reference tracer is evolving in a coherent a jet. Note that
the escaping of the test particles does not mean that the tracer is
not still trapped within the regular jet. It is also possible to dynamically
chose specifically among possible different types of jets using the
averaged speed $\sigma _{L}/\sigma _{D}$ within the jet (see \cite{LZ02}). 

\begin{figure}
\begin{center}\includegraphics[  width=8cm]{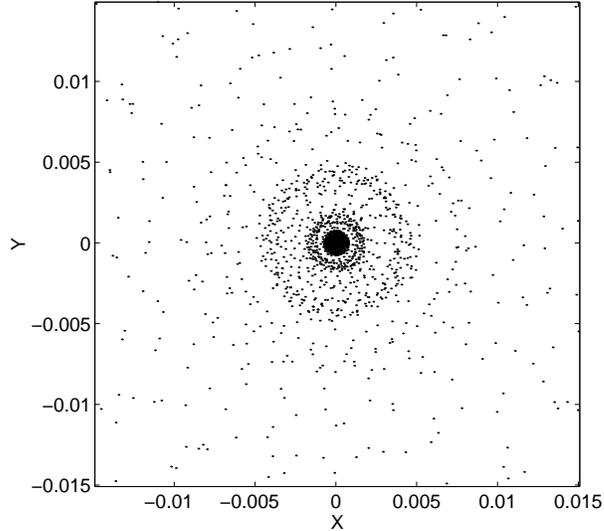}\end{center}

\caption{Structure of a jet in the flow generated by four vortices. The distribution
is not uniform and exhibits a nested set of jets with increasing radii.
\label{Figslowjet}}
\end{figure}
 Having also access to the position of the test particles it becomes
possible to gather some information about the inner structure of the
jet while. The structure is illustrated by plotting the relative position
of a test particles in the frame moving with the reference tracer
in Fig. \ref{Figslowjet}. We can effectively see a structure within
the jet, and a hierarchy of circular (tubular) jets within jets emerge.
This structure is somewhat robust as the same type of structure is
observed when considering a system with 16 vortices \cite{LZ02}.
Note that test particles are coming very close to the reference trajectory
and that this is not an artifact of having initially placed the ghost
in the vicinity of the tracer. Indeed the jet is detected only when
$\sigma _{D}*$ is reached, and for this particular plot the test
particle is not in the close vicinity of the reference tracer at this
moment; the influence of the parameter $\delta $ becomes in this
regards less relevant. This hierarchical structure is reminiscent
of the discrete renormalization group, and we can speculate that log-periodic
oscillation described in \cite{Benkadda99} may be observed. Moreover,
the hierachy allows to clear out the eventual influence of the parameter
$\epsilon $. 

In this paper we have reviewed in details the notion of chaotic jet.
This notion allows to define a proper diagnostic which was able to
capture the singular dynamics responsible for anomalous transport
in systems of point vortices. The possibility to actually visualize
the sturcture of the jet allows to discard the qualitative influence
of the parameters $\epsilon $ and $\delta $, making jets an actual
robust feature of the system dynamics realted to some kind of {}``hidden
order''. In these systems, the jets were located on the boundaries
of coherent structures. We therefore speculate that this behavior
is generic and that the detection of jets should lead to the localization
of the so-called coherent structure responsible for anomalous transport
in more complex systems. The detection of jets in systems exhibiting
typical properties of two-dimensional turbulence is currently under
investigation. Finally we emphasize that the diagnostic we set up
in order to detect coherent jets can be understood as a particular
case of measurements of space-time complexity\cite{Afraimovich03}.

\vspace{0.2cm}

 {\sl Acknowledgements.}

X. Leoncini would like to thank O. Agullo and S. Benkadda for fruitfull
discussions.

\bibliographystyle{elsart-num.bst}
\bibliography{./transport}

\end{document}